\documentclass[symmetry,article,accept,pdftex,moreauthors]{Definitions/mdpi}

\usepackage{amsthm}

\usepackage{wasysym}

\firstpage{1} 
\pubvolume{15}
\issuenum{11}
\articlenumber{1966}
\pubyear{2023}
\copyrightyear{2023}
\externaleditor{Academic Editors: Abraham A. Ungar and M. D. Rodriguez Frias}
\datereceived{27 September 2023} 
\daterevised{13 October 2023} 
\dateaccepted{20 October 2023} 
\datepublished{24 October 2023} 
\hreflink{https://\linebreak doi.org/10.3390/sym15111966} 

\Title{Action Principle for Scale Invariance and Applications (Part I)}

\TitleCitation{Action Principle for Scale Invariance and Applications (Part I)}

\Author{{Andre} 
 Maeder $^{1}$\orcidA{} and Vesselin G. Gueorguiev $^{2,3,}$*\orcidV{}  }  

\AuthorNames{Andre Maeder  and Vesselin G. Gueorguiev}

\AuthorCitation{Maeder, A.; Gueorguiev, V.G.}

\address{%
$^{1}$ \quad Geneva Observatory, University of Geneva, Chemin des Maillettes 51, CH-1290 Sauverny, Switzerland; {andre.maeder@unige.ch} 
\\
$^{2}$ \quad Ronin Institute for Independent Scholarship, 127 Haddon Pl., Montclair, NJ 07043, USA\\
$^{3}$ \quad Institute for Advanced Physical Studies, 1784 Sofia, Bulgaria

}

\corres{Correspondence: {vesselin@mailaps.org} 
} 

\abstract{
On  the  basis of a general action principle, 
we {revisit} the scale invariant field equation using the cotensor relations by Dirac (1973). 
This  action principle also leads to an expression for the scale factor $\lambda$,
which corresponds to the one derived from the gauging condition,
which assumes that a macroscopic empty space  is scale-invariant,  homogeneous, and isotropic.
These results strengthen the basis of the scale-invariant vacuum (SIV) paradigm. 
From the field and geodesic equations, we derive, in current time units (years, seconds), 
the Newton-like equation, the equations of the two-body problem, and its secular variations.  
In a two-body system, orbits very slightly expand, while the orbital velocity  keeps constant during expansion.
Interestingly enough, {Kepler's third} law is a  remarkable scale-invariant  property.
}

\keyword{action principle; scale invariance; scale invariant field equations; geodesic equations; 
Dirac co-calculus; Weyl integrable geometry; scale-invariant cosmology; Kepler's third law} 

\begin{document}
\tableofcontents 

\section{Introduction} \label{intro}

The  scale-invariant vacuum (SIV) paradigm aims to respond to a  fundamental principle expressed by
 \citet{Dirac73}: ``It appears as one of the fundamental principles
in Nature that the equations expressing basic laws should be invariant
under the widest possible group of transformations''. 
Our objective is  to explore whether, in addition to Galilean, Lorentz invariance and  general covariance, some effects of
scale invariance are also present in our   low-density Universe.
This question is fully justified,  since scale invariance is present in Maxwell's equations and general relativity (GR)
in empty space without cosmological constants, charges, and currents.

The presence of matter tends to kill  scale invariance \citep{Feynman63}.
Thus, the question arises of how much matter in the Universe is necessary to suppress scale invariance.
Would one single atom in the Universe be enough to kill scale invariance?
The question of the quantum particle content and the corresponding conformal energy--momentum tensor 
that may arise from a conformal scalar field metric transformation was first studied by \citet{Parker'73}.
A key result of this study was that there is no gravitationally induced particle creation in conformally flat space-times
when the mass of the scalar field can be neglected in the conformal wave equation.
Here, we develop a study of the SIV theory in two stages, 
first by imposing an extremum of the variations of  the action for the scale factor $\lambda$ 
to derive the general scale-invariant equations, and 
second by considering a scale-invariant vacuum state of the Universe  as  a conformally  flat  space-time background  
with zero curvature R and zero associated mass term, to arrive at a specific  SIV expression for  the scale factor
$\lambda(t)=t_0/t$. The two stages are also consistent with the case of non-zero mass, where 
the general equations factor into GR equations without a cosmological constant term and 
equation determining $\lambda$ that absorbs any non-zero cosmological constant term (cf. Section \ref{fixing}).
{Our} study on the causal connexions {in models of expanding universes} 
indicates that scale invariance is certainly  forbidden in cosmological models with  densities above the critical density
$\varrho_{\mathrm{c}}=3 H^2_0/(8 \pi G)$ \citep{MaedGueor21}. This result is in agreement with
the equations of scale-invariant cosmological models, since they show the absence
of  possible expansion solutions  for $\Omega_{\mathrm{m}} \geq 1$
\citep{Maeder17a}.

For universe models with $\Omega_{\mathrm{m}} < 1$, the question remains open, since  scale-invariant
cosmological models do  have solutions (rather close to the $\Lambda$CDM).
In the range of  possible  $\Omega_{\mathrm{m}}$ between 0 and 1,
the higher the density, the smaller the effects from scale invariance. For
$\Omega_{\mathrm{m}} = 0.2$  to 0.3, these limited effects are nevertheless sufficient to drive 
a significant acceleration of the expansion of the universe. 
Of course, the observations will  decide whether the  effects of scale
invariance are effectively present or not. 

Indeed, there is little hope of convincing theoretical   astrophysicists about new developments in gravitation theory if
these are not resting on a well-established action principle. In order to try to  fulfill this requirement,  we  
{revisit} the known results and aim at a  detailed demonstration 
of  the scale-invariant field equation from  an action principle in the line of former developments 
of cotensor calculus and  action  by \citet{Dirac73}.
{For a more modern treatment of the scale-invariant gravity idea see \cite{2018GReGr..50...80W},
which is based on Cartan's formalism, with plenty of practice with differential-forms, 
and along a more traditional abstract scalar field approach, 
which due to its abstractness seems to have stayed disconnected from observational tests.
Here, our approach is more traditional, physically motivated, and with as little general abstraction as possible}.
The interest in such an undertaking is first to obtain a complete derivation of the scale-invariant field
equation by imposing an extremum of the action for small changes $\delta g_{\mu \nu}$. Second, we also
explore the consequences of  an extremum in the action for small variations  $\delta \lambda$ of the
scale factor.   Interestingly enough,  in this  case, the action principle
leads  to a well-defined form of the scale factor $\lambda$,  corresponding to the  gauging condition
\citep{Maeder17a} based on the statement
that the macroscopic empty space should be scale-invariant. This gauging condition replaces the one based on the
``large number hypothesis'' originally proposed by Dirac and also used by {Canuto and his collaborators}
to express $\lambda$ in the scale-invariant framework \citep{Dirac73,Dirac74,Canuto77}.

Several positive results for the SIV theory have already  been obtained: on basic cosmological tests 
\citep{Maeder17a}, 
  on the growth of density fluctuations without the need of dark matter \citep{MaedGueor19}, and 
  on the clustering of galaxies  and galactic rotation \cite{Maeder17c}. 
  In particular, MOND \citep{Milgrom83,Milgrom09} was shown to be a good  approximation of SIV theory
when the scale factor is taken as a constant. Such an approximation applies with an  accuracy better 
than 1 \% for dynamical timescales shorter than about 400 Myr \citep{Maeder23}, 
encompassing the typical rotation time of spiral galaxies.

After a brief recall of some basics of the cotensor calculus, Section \ref{sec2} gives  the action principle  and
a resulting demonstration of the scale-invariant field equations;  the detailed steps  are given in Appendix \ref{appa}.
Section \ref{fixing} shows that the action principle also leads to an  expression  of the scale factor identical to that derived
from  our gauging condition \citep{Maeder17a}.
In Section \ref{sec4}, we examine some consequences of the scale-invariant cosmological equations, 
 {Newtonian} approximation, the two body-problem, and  
{Kepler's third} law.
Some of these properties were previously obtained    in a timescale  suited for  cosmological models; here, we give 
them in a {form more appropriate for} observational studies with current time units  (years or seconds).
Section \ref{sec5} contains the conclusions.

\section{The Scale-Invariant Field Equation}
\label{sec2}
The Einstein  field equation of general relativity (GR) can be derived from
the extremum of an action $ I = I_{\mathrm{G}} +  I_{\mathrm{M}}$ containing a gravitational and a matter
term \citep{Landau66,Weinberg78},~where
\begin{equation}
I_{\mathrm{G}} \, =
\left[\frac{c^4}{16 \pi G}\right] \int \sqrt{-g} \, R(x) \, d^4 x \,,
\label{IG}
\end{equation}
with $R$ being  the curvature scalar. Often, the  multiplicative constants in the bracket are omitted
through choosing an appropriate units system, and we do so below; 
however, they are needed after Equation (\ref{IM}), when we define the energy--momentum tensor. 
The stationarity of the action $I$ classically leads to Einstein theory. As usual, 
{the functional integrand is determined up to a total derivative of a smooth function, 
which often is required to vanish at the integration boundary; that is, often at infinity}.
Terms in $R^2$, $R^3$, etc.  can be added to $R$. Such extensions lead,
in particular, to the family of $f(R)$ theories \citep{Capozziello08,Capozziello11}.

\subsection{Definitions: Weyl Integrable Geometry and Dirac Co-Tensors} \label{defin}

The so-called Weyl integrable geometry (WIG) \citep{Dirac73,Canuto77,BouvierM78} is a particular case of
Weyl's geometry (WG) \citep{Weyl23}, which was initially developed  to express electromagnetism using  a space-time property.  WIG now  forms  an appropriate framework
for studying scale invariance. Weyl's geometry (WG), 
{just like in GR}, is endowed with a quadratic form $ds^2=g_{\mu\nu}dx^{\mu} dx^{\nu}$.
In addition,   the length $\ell$ of a vector  $a^{\mu}$ is determined by a scale factor $\lambda(x^{\mu})$
and this also applies to the line element $ds$,
\begin{equation}
\ell^2 \, = \, \lambda^2 (x^{\mu}) \, g_{\mu\nu} \, a^{\mu} a^{\nu}\,,  \quad \mathrm{and}
\; \; ds'=\lambda (x^{\mu})  \, ds\,.
\label{ell}
\end{equation}

This last equation relates the line element  $ds$  in WG space to $ds'$ in another space
(here, $ds$ is in a WG space; while $ds'$ is in Einstein GR space, which is a particular WIG space).
Thus, the two {considered} spaces (GR and WIG)  are conformally  equivalent
using the metric conformal transformation  $g'_{\mu \nu} = \lambda^2 g_{\mu \nu}$.
Note that one has to distinguish between scale coordinate transformations, conformal coordinate transformations,
and conformal transformations of the metric \cite{Parker'73}.
Usually, the primed quantities will be in {a GR} frame.

A key property  concerns the  transport of a vector  from a point
$P_1(x^{\mu})$ to a nearby point  $P_2(x^{\mu} +dx^{\mu} )$. During such a transport, the  length of the vector
changes as follows:
\begin{equation}
d\ell \, = \, \ell \, \kappa_{\nu} \, dx^{\nu}.
\label{delu}
\end{equation}

Here, $\kappa_{\nu}$ is called  the coefficient of metrical connection.
In GR,  the coefficient of metrical connection is
$\kappa'_{\nu}=0$. 
The specificity of WIG \citep{Canuto77},
with respect to the classical WG~\citep{Weyl23}, is that the metrical connection  $\kappa_{\nu}$   is the gradient  of a scalar field  $\varphi$,
namely
\begin{equation}
\kappa_{\nu}= -\varphi_{, \nu} \quad  \mathrm{with} \ \; \varphi  =  \ln \lambda,
\quad i.e., \; \kappa_{\nu} = - \frac{\partial \ln \lambda}{\partial x^{\nu} }.
\end{equation}

This implies that  $\partial_{\nu} \kappa_{\mu}=\partial_{\mu} \kappa_{\nu}$.
Thus, the parallel displacement of a vector along a closed loop in WIG does not change
its length; as a consequence,  the length change of a vector does not depend on the path followed.
Weyl's original theory was withdrawn  due to a criticism by \citet{Einstein18}, 
who pointed  out that the properties of  atoms  would then depend on their past world lines. 
Thus, the  atoms of an element in an electromagnetic field could not show sharp lines.

The general covariance of GR {requires} tensorial expressions. Scale invariance
demands further developments: the cotensors. Many mathematical tools for Weyl's geometry \citep{Weyl23}
have been  applied and  developed  by \citep{Eddington23,Dirac73}.
A quantity $Y,$ scalar, vector, or tensor, 
from which one obtains a scale-invariant GR object   $Y'$, and  
which upon a scale transformation changes  like $Y\rightarrow\,Y' \,= \, \lambda^n(x) \, Y$, 
is said to be a {\emph{{coscalar, covector, or cotensor}
 }} of  power $\Pi(Y) =n$; one speaks of {\emph{{scale covariance}}}.
For $n=0$ ({\emph{{scale invariance}}}), one has an {\emph{{inscalar, invector, or intensor}}}.
The ordinary derivative  $Y_{,\mu}$ and the covariant derivative $Y_{;\mu}$   are not necessarily
{\emph{{co-covariant}}}  (a definition by Dirac) or invariant,
but  derivatives  with such properties can be defined. As an example,
let us perform a derivative and a change of  scale for  scalar $S$; using (\ref{delu}) we have
\begin{eqnarray}
S'_{ ,\mu}=  (\lambda^n  S)_{ \mu}= \lambda^n  S_{\mu}+ n \lambda^{n-1} \lambda_{\mu} S=
\lambda^n   (S_{\mu} - n \kappa _{\mu} S)   \, .
\label{fnu}
\end{eqnarray}

The co-covariant derivative $ S_{*\mu}$ of a coscalar $S$ is thus
\begin{equation}
S_{*\mu}=  S_{\mu} - n \kappa _{\mu} S \, ,
\label{scal}
\end{equation}
which is a covector   of power $n$
(ordinary derivatives are denoted by  $S_{,\mu}$ or  just by $S_{\mu}$ when there is no obviously possible confusion).
Co-covariant derivatives   preserve scale covariance with the same power.
First and second co-covariant derivatives of vectors and tensors can also be defined.
Operations on covectors and cotensors can  be performed.
A brief summary
of cotensor calculus was given by \citet{Canuto77}. Many useful
expressions can also be found in \citet{Dirac73}, as well as in \citet{BouvierM78} for geodesics,
isometries, and killing vectors. (We limit the presentation here to what is really needed.)

\subsection{The Scale-Invariant Action Principle}

We note that in current
scalar--tensor  theories of gravitation, often aimed at studying scale invariance,  a new field $\varphi$
determined by the specific aims of the proponents is usually introduced in  the action,
in addition to the choice for $\lambda$; see for example \citet{Clifton12} and \citet{FerreiraT20}, as well as \citet{Parker'73}.
This field may be a multiplier of the curvature scalar $R$ \citep{Weinberg78},
of   functions and  derivatives  of $\varphi$.
In WIG, we identify  the field $\varphi$  with the conformal scale factor $\lambda$ of Equations (\ref{ell}) and (\ref{delu}),
thus $\varphi =  \lambda$  ($\lambda$  being of power $\Pi(\lambda)= -1$
since $\lambda'=1=\lambda^{-1} \lambda$).
Since the  conformal scale factor $\lambda$ appears in the covariant derivatives, 
it is not expected to have a particular particle content in the same sense that 
the  Levi--Civita connection coefficients are not expected to give rise to a particular particle content.
Thus, no new field is introduced, and the theory, having no
additional degree of freedom, is indeed much more constrained than current scalar--tensor theories.
The SIV theory is a particular case of scalar-theories, or rather
``a cotensor theory``, being based on  cotensoral expressions,  satisfying  both covariance and scale invariance.

The corresponding curvature scalar  $^*R$ is of power $\Pi(*R)= -2$  
(meaning that $^*R $ behaves like $\lambda^2$). 
In order to make the action  $ I_{\mathrm{G}}$ an inscalar,  we take
$\lambda^{2}  \, ^*R$  (power -4)  in  the expression of the action, since $\sqrt{-g}$ is of power 4.
 {Just like} scalar--tensor theories, the action contains derivatives of the field; of course, only terms of quartic power are possible,
due to the necessary global invariance.  This can lead to  two additional possible terms of power $\Pi= -4$
multiplied by constants $c_1$ and $c_2$ in the action (\ref{ig}). The action writes \citep{Dirac73},
\begin{equation}
\delta I_{\mathrm{G}}\, = \,\delta   \int  \, \left( -\lambda^2 \,^{*} R(x) +
c_1 \, \lambda^{*\mu} {\lambda_{*\mu}}+ c_2 \, \lambda^4  \right) \sqrt{-g} \, d^4 x \,,
\label{ig}
\end{equation}
where the curvature scalar $^{*}R$  is  \citep{Dirac73},
\begin{equation}
^{*}R \, =\,^{*}R^{\mu}_{\mu} \, = 
R-6\kappa^{\mu}_{  ; \mu} +6 \kappa^{\mu} \, \kappa_{\mu},
\label{rstar}
\end{equation}
\noindent
where  $R$ refers to the curvature scalar of GR. 
The symbol ``;'' expresses the usual covariant derivatives related to $g_{\mu\nu}$.
This expression was given in Equation (89.2) by \citet{Eddington23} and later by \citet{Parker'73},
where one has to map $\Omega$ to $\lambda$. {{In} 
 order to match the expression derived by \citet{Parker'73} with the original expression by \citet{Dirac73},
one has to use the conformally transformed metric $\tilde{g}_{\mu\nu}$ to  {raise}
and lower indexes for the $*$ objects on the left hand side,
while in the  second equality using covariant derivatives `$;$' based on $g_{\mu\nu}$ which
corresponds to the Parker view point.
That is,
 $\,^{*}R^{\mu}_{\mu} \, = \lambda^{-2}g^{\mu\nu}\;R_{\mu\nu}(\tilde{g})\, =
R-6\kappa^{\mu}_{  ; \mu} +6 \kappa^{\mu} \, \kappa_{\mu}.$ 
This way one recovers Equation (25) in Parker's paper upon identifying  $\lambda$ with $\Omega$ 
and moving  $\Omega$ to the r.h.s of the last equality}).
As such, the above action  applies to the vacuum,
since the matter contribution is not yet accounted for.
The action should be stationary for arbitrary small variations of $g_{\mu\nu}$ and of $\lambda$.

In the expression $\lambda^2\, {^*R}$, the second term on the right hand side (r.h.s.), 
i.e., $\lambda^2 \kappa^{\mu}_{  ; \mu}$, can be related to 
\begin{equation}
(\lambda^2 \kappa^{\mu})_{  ; \mu} = \lambda^2  \kappa^{\mu}_{  ; \mu}+2 \lambda \kappa^{\mu}  \lambda_{\mu} \, .
\end{equation}

We also note that, thanks to Equation (\ref{scal}) with $n=-1$,
\begin{equation}
{\lambda^{*\mu}} {\lambda_{*\mu}} =g^{\mu\nu}\lambda_{*\mu}\lambda_{*\nu} 
={ \lambda^{\mu}} \lambda_{\mu}+ {\lambda^2} { \kappa^{\mu} } \kappa_{\mu}
+2 \lambda {\kappa^{\mu}} {\lambda_{\mu}}\,  .
\end{equation}

One can now look at the first two terms in the action integral:
\begin{eqnarray}
-\lambda^2 {^{*}R}+  c_1 \, \lambda^{*\mu} {\lambda_{*\mu}}= -\lambda^2  R-6 \lambda^2 \kappa^{\mu} \kappa_{\mu}
+ 6 (\lambda^2 \kappa^{\mu})_{ ; \mu}    \\  \nonumber
-12 \lambda  \kappa^{\mu} \lambda_{\mu}  +  c_1 \lambda^{\mu} \lambda_\mu+
c_1 \lambda^2  \kappa^{\mu} \kappa_{\mu}+ 2 c_1 \lambda \kappa^{\mu} \lambda_{\mu} \,.
\end{eqnarray}

We choose the  {constant} $c_1$ to be $c_1=6$,
in order to reproduce the usual kinetic term for  a scalar field $\varphi$ related 
to small deviations near $\lambda=1$; that is, $\lambda\approx1+\varphi$;
to be touched upon in the next section.
Then, the above expression simplifies to
\begin{equation}
-\lambda^2  {^{*}R}+  6 \, \lambda^{*\mu} {\lambda_{*\mu}}= -\lambda^2  R+ 6 (\lambda^2 {\kappa^{\mu}}){_{ ; \mu}}
+  6 \lambda^{\mu} \lambda_\mu\,.
\label{ez}
\end{equation}

We note  that $(\lambda^2 \kappa^{\mu}){_{ ; \mu}} \sqrt{-g}= (\lambda^2 \kappa^{\mu} \sqrt{-g})_{,\mu}$
(since $\Gamma^{\alpha}_{\alpha \mu} = \frac{\partial  \ln \sqrt{-g}}{\partial x^{\mu}}$). 

This is an exact differential and may be eliminated from the action integral \citep{Landau60}. 
The action is then \cite{Parker'73}:
\vspace{-6pt}
\begin{equation}
\delta I_{\mathrm{G}}\, = \,\delta   \int  \, ( -\lambda^2 \, R  + 6 \, \lambda^{\mu} \lambda_{\mu}
+ c_2 \, \lambda^4  ) \sqrt{-g} \, d^4 x \,.
\label{ig2}
\end{equation}

The variations of the action can be studied with respect to  small variations of both $\delta g_{\mu \nu}$ and $\delta \lambda$.
The details of the calculations are given in the Appendix.
The extremum with respect to  $\delta g_{\mu \nu}$  leads to the expression
of the  field equation below  with the properties of covariance, as in GR   with scale invariance in addition.

\subsection{The Scale-Covariant Field Equation}   \label{geneq}

The development of the scale-covariant  field equation in a vacuum  with  a cosmological constant
is presented  in the Appendix, it gives Equation (\ref{f6x}),
\begin{eqnarray}
R^{\mu \nu} - \frac{1}{2} g^{\mu \nu} R - 2 g^{\mu \nu} \frac{(\lambda^{\rho}){_{; \rho}}}{\lambda}
+ 2 \frac{ ( \lambda^{\mu}){^{; \nu}}}{\lambda} \nonumber \\
+g^{\mu \nu} \frac{\lambda^{\alpha} \lambda_{\alpha}}{\lambda^2}   - 4 \frac{\lambda^{\mu} \lambda^{\nu}}{\lambda^2}+
\lambda^2 \Lambda_{\mathrm{E}} \,g^{\mu \nu} =0 \, .
\label{f6xx}
\end{eqnarray}

We  perform the following identification $ c_2  = 2\Lambda_{\mathrm{E}}$, and   $\Lambda =\lambda^2 \Lambda_{\mathrm{E}}$,
where $\Lambda_{\mathrm{E}}$ is the cosmological constant of GR (not necessarily that of the Einstein static universe)
and $\Lambda$  the corresponding cosmological constant in WIG space.
(The notation $\Lambda_{\mathrm{E}}$ is used  to avoid any confusion  with  $\Lambda$ in SIV).
The above equation applies in general; thus, apart from $\Lambda$, it  forms  the first member
of the   field equation when matter-energy is present.
Aside from the {absent} mass term $g_{\mu\nu}\mu^2_0\lambda^2$, the $\lambda$ dependent part of 
the above expression (\ref{f6xx}) is practically the same modified energy-momentum tensor as discussed by \citet{Parker'73}. 
Therefore, if there is any particle content related to $\lambda$,
it can be viewed as a energy--momentum tensor of such matter that causes the gravitational geometry of the spacetime, 
in accord with the Einstein GR view point.

If one desires, then the integration measure in (\ref{ig2}) could be set to be $\lambda^2\sqrt{-g}d^4x$, 
resulting in the usual Hilbert--Einstein Lagrangian density for gravity ${\cal{L}}_G=R$, 
while the remaining terms can be identified with the matter Lagrangian  
${\cal{L}}_M\sim\varphi^\mu\varphi_\mu+\tilde{c}_2\,e^{2\varphi}$, where $\varphi=\log{\lambda}$.
Such a treatment results in a stress--energy tensor $\Lambda_{\mu\nu}=\frac{2}{\sqrt{-g}}\frac{\delta{I_M}}{\delta{g^{\mu\nu}}}$ 
of power  $\Pi(\Lambda_{\mu\nu})=-2$ as derived by \citet{Parker'73}. 
{This is consistent with our construction of the overall action as a conformally invariant object}. 
Thus,  from the scaling of $\Pi(\sqrt{-g}=+4),\;\text{and} \; \Pi(g^{\mu \nu})=-2$
one obtains  $\Pi(\Lambda_{\mu\nu})=-4+2=-2$ coming from the denominator in  $\Lambda_{\mu\nu}$. 

If there is any additional matter, then there will be a general matter action $I_M=\int{{\cal{L}}_M}\sqrt{-g}d^4x$,
which via standard considerations would result in  the stress--energy tensor 
$T_{\mu\nu}=\frac{2}{\sqrt{-g}}\frac{\delta{\sqrt{-g}{\cal{L}}_M}}{\delta{g^{\mu\nu}}}$
or simply $T_{\mu\nu}=\frac{2}{\sqrt{-g}}\frac{\delta{I_M}}{\delta{g^{\mu\nu}}}$.
If the matter action is reparametrization-invariant, as needed to understand why 
{there are} only electromagnetic and gravitational classical long-range forces \cite{sym13030379}, 
then one would again obtain  $\Pi(T_{\mu\nu})=-2$. 
As the name suggests, reparametrization invariance is related to the freedom of choosing any  
reasonable parametrization for a process under consideration. In particular, this could be accomplished 
via $\lambda$ being only time-dependent but not space-dependent.

This means that there is a large class of matter models that can result in a stress--energy tensor 
that is of power $\Pi(T_{\mu\nu})=-2$. Since, $\Pi(\lambda)=-1$, then one can construct a scale-invariant 
stress--energy tensor of zero power by considering $T'_{\mu\nu}=\lambda^{-2}T_{\mu\nu}$.

Therefore, in the presence of  matter, the variation $\delta I_{\mathrm{M}}$ of  the  matter action
can be considered to take the form:
\color{black}
\begin{equation}
\delta  I_{\mathrm{M}}=  
-\left[\frac{1}{2} \right] \int \, T_{\mu \nu}\, \delta g^{\mu \nu}  \lambda^2 \sqrt{-g}  d^4 x  \,,
\label{IM}
\end{equation}
which contains  scale-invariant energy--momentum tensor  $T_{\mu \nu}$. 
Equation (\ref{IM}) is consistent with the scale invariance of $T_{\mu \nu}$,
since $\Pi(\lambda^2) = -2, \; \Pi(\sqrt{-g}=+4), \; \Pi(g^{\mu \nu})=-2$.

The scale invariance of  the momentum--energy tensor  has some implications for  densities and pressures, if we consider, for example,
the typical case  of a perfect fluid \citep{Canuto77},
\begin{equation}
T_{\mu \nu}=T '_{\mu \nu} \Rightarrow
( p+\varrho) u_{\mu} u_{\nu} -g_{\mu \nu } p =
( p'+\varrho') u'_{\mu} u'_{\nu} -g'_{\mu \nu } p' \, .
\label{pr2}
\end{equation}

There, the  velocities $u^{\mu}$ and $u'_{\mu}$ transform like
\begin{eqnarray}
u'^{\mu}&=&\frac{dx^{\mu}}{ds'}=\lambda^{-1} \frac{dx^{\mu}}{ds}= \lambda^{-1} u^{\mu} \, , \nonumber \\
\mathrm{and} \; \;
u'_{\mu}&=&g'_{\mu \nu} u'^{\nu}=\lambda^2 g_{\mu \nu} \lambda^{-1} u^{\nu} = \lambda \, u_{\mu} \, .
\label{pl1}
\end{eqnarray}

The contravariant and covariant components of a vector have different powers, and their covariant
derivatives are also  different.
The energy--momentum tensor scales like
\begin{equation}
( p+\varrho) u_{\mu} u_{\nu} -g_{\mu \nu } p =
( p'+\varrho') \lambda^{2} u_{\mu} u_{\nu} - \lambda^2 g_{\mu \nu } p' \, ,
\end{equation}
with implications for  $p$ and $\varrho$ \citep{Canuto77},
\begin{equation}
p = p' \, \lambda^2 \, \quad \mathrm{and} \quad \varrho = \varrho' \, \lambda^2 \, .
\label{ro2}
\end{equation}

Thus, the invariance of the stress--energy tensor $T_{\mu \nu}$ implies that the 
pressure and density are  coscalars of power $\Pi(p)=\Pi(\rho)=-2$.

We  now express the sum
$\delta I= \delta  I_{\mathrm{G}}+\delta  I_{\mathrm{M}}=0$, see  Equations (\ref{IG}) and (\ref{IM}),
\begin{eqnarray}
\delta I =  \left[\frac{c^4}{16 \pi G}\right] \int d^4x\sqrt{-g}\lambda^2 \left\{ 
R^{\mu \nu} -\frac{1}{2} R  g^{\mu \nu}
- 2 g^{\mu \nu} \frac{(\lambda^{\rho}){_{; \rho}}}{\lambda}
+ 2 \frac{ ( \lambda^{\mu}){^{; \nu}}}{\lambda} \nonumber \right. \\ \left.
+g^{\mu \nu} \frac{\lambda^{\alpha} \lambda_{\alpha}}{\lambda^2}   - 4 \frac{\lambda^{\mu} \lambda^{\nu}}{\lambda^2}+
\lambda^2 \Lambda_{\mathrm{E}} g^{\mu \nu}
+\left[ \frac{8\pi G}{c^4} \right] T^{\mu \nu}  \right\} \delta g_{\mu \nu}.
\label{final}
\end{eqnarray}

We have used the fact that,  for any tensor $A^{\alpha \beta}$, one has $A^{\alpha \beta} dg_{\alpha \beta}=
- A_{\alpha \beta} dg^{\alpha \beta}$.
The constant terms are indicated in brackets. Thus, in the final equation
the energy--momentum tensor  is  preceded by the  constant  $[ \frac{8\pi G}{c^4} ]$.
It is also scale-invariant as the vacuum contribution. 
The constant of gravity $G$ is here kept as a true constant.

We note that $\Lambda$ is  a coscalar of power $\Pi(\Lambda)=-2$,
consistently with the  previous results for $p$ and $\varrho$. 
This {correspondence ensures} the scale invariance
of the second member of the field equation,
\begin{eqnarray}
R^{\mu \nu} - \frac{1}{2} g^{\mu \nu} R - 2 g^{\mu \nu} \frac{(\lambda^{\rho}){_{; \rho}}}{\lambda}
+ 2 \frac{ ( \lambda^{\mu}){^{; \nu}}}{\lambda}
+g^{\mu \nu} \frac{\lambda^{\alpha} \lambda_{\alpha}}{\lambda^2}
- 4 \frac{\lambda^{\mu} \lambda^{\nu}}{\lambda^2} = \nonumber\\
- \frac{8 \pi G}{c^4} T^{\mu \nu} - \Lambda  g^{\mu \nu}  \, .
\label{f67}
\end{eqnarray}

Hereafter, following the general practice, we use  $c=1$.
In the above equation, the
first member is written in {terms} of derivatives of $\lambda$, we can also write it in terms of
of $\kappa_{\nu} =  - \lambda_{\nu}/\lambda$, noting, for example, that
$\kappa_{\nu ; \mu}= - \frac{\lambda_{\nu ; \mu}}{\lambda} + \frac{ \lambda_{\nu} \lambda_{\mu}}{\lambda^2}$.
This gives, for example,  in the covariant form,
\begin{eqnarray}
R_{\mu \nu} - \frac{1}{2} \ g_{\mu \nu} R-\kappa_{\mu ;\nu}-\kappa_{ \nu ;\mu}
-2 \kappa_{\mu} \kappa_ {\nu}
+ 2 g_{\mu \nu} \kappa^{ \alpha}_{;\alpha}
- g_{\mu \nu}\kappa^{ \alpha} \kappa_{ \alpha} = \nonumber  \\
-8 \pi G T_{\mu  \nu} - \Lambda \, g_{\mu \nu}. \quad \label{field}
\end{eqnarray}

This equation is similar to  the one   obtained through development of the Ricci tensor
in  the cotensor  calculus, see \citet{Canuto77}. There, $R_{\mu \nu}$ and $R$ are the usual terms of GR.
This equation is the fundamental equation of the gravitational field, with the properties of both  general covariance
and scale invariance. The terms of the action with variations  in $\delta \lambda$  
are discussed below  in relation with the 
``gauge fixing'' that determines a particular functional form of $\lambda$, 
based on a physical assumption about the vacuum.
Notice that this is a physically justified result and therefore it is not a gauge choice  
for a standard gauge symmetry model. Usually, any gauge choice would result in the same
description of a physical system and, in this respect, the gauge choice is irrelevant, besides allowing us to perform the computation more {easily}.
In the SIV paradigm, the choice of  $\lambda$ has physical implications that are manifested
in the equation of motion for the matter particles, as seen by the presence of $\kappa$-terms in (\ref{geod}) {below}.

\section{The Action Principle and the SIV Gauge}  \label{fixing}

To apply the field equation of GR,  it is necessary to specify the form of the metric, which defines the geometrical properties. 
In SIV theory,  an additional  condition is needed  to fix the functional form of $\lambda$.
\citet{Dirac73} and \citet{Canuto77} chose the so-called ``large number hypothesis''.
The related numerical coincidences have received
a variety of  interpretations, among which is the anthropic one  \citep{Weinberg89}.
{Our} choice is to adopt  the  following condition for the physical gauge \citep{Maeder17a}:
``The macroscopic empty space is scale invariant, homogeneous, and isotropic''.
The equation $p= - \varrho c^2$ for the vacuum  allows the medium density
to remain constant for an adiabatic expansion or contraction \citep{Carroll92}, which  implies  that changes in the spatial
scales of the empty space do not modify its properties.  Thus, consistently,  the  only possible dependence, if any, of the scale factor
$\lambda$ is a dependence on  time.

\subsection{The Scale-Invariant Vacuum (SIV) Gauge for $\lambda$} \label{SIVgauge}

Above,   the extremum of the action  was considered with respect to small  variations of $\delta g_{\mu \nu}$, 
we can also consider the extremum with respect to variations of $\lambda$.
Collecting  the terms  with $\delta \lambda$ in Equations (\ref{e3}), (\ref{f4}) and (\ref{f5})
for the vacuum contributions to the action integral, we obtain as a condition for a stationary  action
\begin{equation}
( -12 \lambda ^{\alpha}_{; \alpha}- 2 \lambda R +4  c_2 \lambda^3) \sqrt{-g} \delta \lambda  = 0 \,.
\label{f7}
\end{equation}

Making the identification  $c_2 = 2\Lambda_{\mathrm{E}}$, we obtain
\begin{equation}
6 \lambda ^{\alpha}_{; \alpha}+ \lambda R  =  4 \lambda^3 \, \Lambda_{\mathrm{E}}\,.
\label{f8}
\end{equation}

By identifying $\lambda$ with $\phi$, this is the same as Equation(35) derived by \citet{Parker'73}, 
but here the mass term $\mu^2_0\phi$ is missing.
The curvature scalar $R$ is different from zero, as expected  in de Sitter space-time.
However, the de Sitter space-time  is conformal to the flat Minkowski space; for the following particular condition,
these two spaces are even strictly identical~\citep{Maeder17a},
\begin{equation}
\frac{3 \, \lambda^{-2}}{\Lambda_{\mathrm{E}} \,  t^2} = \,1 \,.
\label{LL}
\end{equation}

\textls[-15]{In this case, since in Minkowski space $R=0$, the action  principle results in the expression}
\begin{equation}
6\,  \lambda ^{\alpha}_{; \alpha}\, =   4 \lambda^3 \, \Lambda_{\mathrm{E}}\, , \quad \mathrm{which \; gives } \; \; \;
6 \, \ddot{\lambda} \, =    4 \lambda^3 \, \Lambda_{\mathrm{E}}\, .
\label{f9}
\end{equation}

The gauging conditions based on the hypothesis of the homogeneity and isotropy of the empty space
{impose} the following two conditions  \citep{Maeder17a}:
\begin{eqnarray}
\  3 \, \frac{ \dot{\lambda}^2}{\lambda^2} \, =\, \lambda^2 \,\Lambda_{\mathrm{E}}  \,
\quad \mathrm{and} \quad 2 \, \frac{\ddot{\lambda}}{\lambda} - \frac{ \dot{\lambda}^2}{\lambda^2} \, =
\, \lambda^2 \,\Lambda_{\mathrm{E}}  \, .
\label{diff1}
\end{eqnarray}

Upon taking the trace of (\ref{field}), one derives the first equation above while using the vacuum assumption $R_{\mu\nu}=T_{\mu\nu}=0$.
We can easily verify that these are equivalent to \mbox{Equation (\ref{f9})}. Introducing the first (\ref{diff1}) into the second one, we obtain
\begin{equation}
2 \, \frac{\ddot{\lambda}}{\lambda} - \frac{1}{3}  \lambda^2 \,\Lambda_{\mathrm{E}} \, =
\, \lambda^2 \,\Lambda_{\mathrm{E}} \quad \mathrm{and\; thus} \; \;
6 \, \ddot{\lambda} \, =    4 \lambda^3 \, \Lambda_{\mathrm{E}}\, .
\label{diff2}
\end{equation}

\textls[-25]{The solution to this differential equation, based on the first equation in (\ref{diff1}) , is very simple:}
\begin{equation}
\lambda \, = \, \sqrt{\frac{3}{\Lambda_{\mathrm{E}}}} \, \frac{1}{t} \,,
\label{condj}
\end{equation}

Noticeably, this last relation is the same as the above Equation (\ref{LL}), which was necessary to assume
the identity of the Minkowski and de Sitter spaces. This verifies
the consistency  of the above expression for $\lambda$.
It is also remarkable that these equations
{ imply $\Lambda_\mathrm{E}=$const, 
consistently with the properties of the Einstein's cosmological constant.
However, there is a deeper connection to the coscalar power of $\Lambda$;
that is, it is a coscalar of power 2, resulting in $\Lambda\lambda^{2}=\Lambda'=\Lambda_\mathrm{E}$.
Equation (\ref{condj}) implies a specific} relation between the current age of the universe and $\Lambda_\mathrm{E}$
when choosing units such that $\lambda_0=1$ in the current~epoch.

Now, we can go back and take another viewpoint  of (\ref{field}), 
this time by adopting the functional form of $\lambda=t_0/t$ (\ref{condj}).
Such a choice will remove all the $\kappa$ terms, along with the $\Lambda$ term, 
and will result in the standard Einstein GR equations $R_{\mu\nu}-g_{\mu\nu}R/2=-8\pi\,GT_{\mu\nu}$
without a cosmological constant; thus, any matter content is possible after removing the 
$\Lambda$ term through the selection of the SIV gauge  (\ref{condj}) for the 
scale factor $\lambda(t)$.

Equations (\ref{f9}) and (\ref{diff1}) also mean that  $\Lambda_{\mathrm{E}}$,
or the energy density of the vacuum (since  $\Lambda_{\mathrm{E}}= 8 \pi G \rho_{\mathrm{vac}}/c^2$),
is a function of the time-variations of $\lambda$; in particular, the first of (\ref{diff1}) illustrates this relation well.
Thus, the energy density of the vacuum  may be expressed as the gradient of a scalar function  $\psi$
\citep{MaedGueor21},
\begin{equation}
\varrho_{\mathrm{vac}} = \frac{1}{2} C \dot{\psi}^2\,, \quad \mathrm{with} \; \; \psi= -\frac{\dot{\lambda}}{\lambda}\, ,
\label{pvac}
\end{equation}
with $C= \frac{3}{4 \pi G}$ { and c = 1 (a reminder to the reader). 
According to \citep{MaedGueor21}, the above $\psi$} may play the role of a ``rolling field'' during the inflation.

In standard cosmological models based on GR (e.g., Friedmann and $\Lambda$CDM models),
the constant $\Lambda_{\mathrm{E}}$
is not  a direct function of the matter content.
Thus, through   the above relations,
the same applies to the form of the scale factor $\lambda \sim 1/t$, 
which {just like} $\Lambda_{\mathrm{E}}$ 
is independent from the matter content in the Universe, generally expressed by $\Omega_{\mathrm{m}}$.
However, the matter content may drastically limit the range of possible variations in the above $t$ parameter (thus, limiting the range
of $\lambda$-variations), but it does not modify  the functional dependence $\lambda(t)$, see Section \ref{constr}. 

For now, we conclude that  the action principle applied to the vacuum leads to the same expression  of the scale factor
as obtained by the abovementioned gauging condition.
This gives  support to  the above  gauging condition, relating   $\lambda$
and $\Lambda_{\mathrm{E}}$,  and to its  significance. We note that there are also positive implications for the well-known
cosmological constant problem \citep{MaedGueor21}.

\subsection{The Geodesics From an Action Principle} \label{geode}

To study  dynamics, we also need an equation of motion.
The equation of geodesics in Weyl's geometry was first derived by \citet{Dirac73}. 
It obeys the following equation:
\begin{equation}
\frac{du^{\alpha}}{ds}+ \Gamma^{\alpha}_{\mu \nu} u^{\mu} u^{\nu} -\kappa_{\mu}u^{\mu} u^{\alpha}+ \kappa^{\alpha} u_{\mu} u^{\mu}= 0 \, ,
\label{geod}
\end{equation}

It is customary to chose the parametrization such that $u_{\mu} u^{\mu}=\pm1$, depending on the signature of the metric used. 
The geodesic of a free particle can  be obtained from the condition that the following action is minimum \citep{BouvierM78},
\begin{equation}
\delta I = \int^{P_2}_{P_1}  \delta(\lambda  \, ds )= \int^{P_2}_{P_1}  \delta ds' =0 .
\end{equation}

This is a one-dimensional problem of an unknown function $x^{\mu}$ of time.
The above condition means, in fact, that the corresponding path is an extremum, 
i.e., in the Riemann space {it is a minimum,} 
while for pseudo-Riemannian space with Lorentzian metric, it is a maximum.
The application of this action principle confirms
the above equation of the~geodesics. 

\section{Basic Applications of the Scale-Invariant Dynamics}
\label{sec4}
{
In this section, we present what we feel is the minimally viable demonstration of the SIV paradigm and the relevant equations. 
After all, this paper is titled the Action Principle for Scale Invariance and its Applications (Part I), which is the first part dedicated to mathematical formalism. We are in the process of preparing an upcoming manuscript (Part II),  where we plan to discuss specific results related to the early Universe;  that is, applications to inflation, Big Bang nucleosynthesis, and the growth of the density fluctuations within the SIV; 
while, in the late time Universe, the applications of the  SIV paradigm are related to scale-invariant dynamics of galaxies, MOND,  dark matter, and the dwarf spheroidals,  where one can find MOND to be a peculiar case of the SIV theory, as well as possible SIV effects in the Earth--Moon system.
These represent about six different applications, which would have made this paper too large for a normal paper, so here we only focus on two key demonstrations of the SIV paradigm: scale invariant cosmology as the initial motivation and the equations of motion as pertaining to  Kepler's third law.}

\subsection{Cosmological Constraints}  \label{constr}

The main consequences  of the field and geodesic equations for the dynamics
are now  drawn in view of future comparisons with observations.
The field equation with the FLWR metric, together  with the gauging conditions (\ref{diff1}),
leads to the cosmological equations \citep{Maeder17a},
\begin{eqnarray}
\frac{8 \, \pi G \varrho }{3} = \frac{k}{a^2}+\frac{\dot{a}^2}{a^2}+ 2 \,\frac{\dot{a} \dot{\lambda}}{a \lambda} \, ,
\label{E1} \\
-8 \, \pi G p  = \frac{k}{a^2}+ 2 \frac{\ddot{a}}{a}+\frac{\dot{a^2}}{a^2}
+ 4 \frac{\dot{a} \dot{\lambda}}{a \lambda}  \,\label{E2} \\
\mathrm{leading \;to} \quad - \frac{4\pi G}{3} \left(3p+\varrho \right) =
\frac{\ddot{a}}{a} + \frac{\dot{a} \dot{\lambda}}{a \lambda} \, , \label{E3} \\
\mathrm{with} \; \;\varrho \, a^{3(1+c^2_s)} \lambda^{1+3c^2_s} = \mathrm{const.} \label{conserv}
\end{eqnarray}

The last expression is the  conservation law,  with a sound velocity $c^2_s =0$
for a dust model and $c^2_s=1/3$ for the radiative era.  If  $\lambda$ is a constant ($\lambda=1$), the derivatives of $\lambda$
vanish  and one is brought back to the Friedmann  equations. 
The extra term in Equations (\ref{E1})--(\ref{E3}) represents  an additional acceleration {\emph{{in the direction of motion}
}}.   
For an expanding Universe, this extra force produces an accelerated expansion.
For  a contraction,  the additional  term   favors collapse, cf.
the growth of density fluctuations \citep{MaedGueor19}.
Analytical solutions  for the flat SIV models with $k=0$   were found
for the  matter  era \citep{Jesus18},
\begin{equation}
a(t) \, = \, \left[\frac{t^3 -\Omega_{\mathrm{m}}}{1 - \Omega_{\mathrm{m}}} \right]^{2/3}\, , \quad \mathrm{thus} \; \;
t_{\mathrm{in}} \, = \, \Omega^{1/3}_{\mathrm{m}}\,.
\label{Jesus}
\end{equation}

These equations  allow flatness for different values of  $\Omega_{\mathrm{m}}$, unlike the  Friedmann models.
The timescale $t$   is $t_0=1$ at present,   with $a(t_0)=1$.
The graphical solutions are relatively  close to the corresponding $\Lambda$CDM models \citep{Maeder17a},
the deviations being larger at very low $\Omega_{\mathrm{m}}$-values.
The initial time for $a( t_{\mathrm{in}})=0$ is $t_{\mathrm{in}}= \Omega^{1/3}_{\mathrm{m}}$;
the dependence in 1/3 produces a  rapid increase in $ t_{\mathrm{in}}$   for increasing  $\Omega_{\mathrm{m}}$.
For $\Omega_{\mathrm{m}}=0, 0.01, 0.1, 0.3, 0.5$, the values of  $ t_{\mathrm{in}}$ are 0, 0.215, 0.464, 0.669, and 0.794, respectively.
This strongly reduces the possible range of $\lambda(t)= (t_0/t)$
(which varies between $1/t_{\mathrm{in}}$ and $1/t_0=1$)  for increasing $\Omega_{\mathrm{m}}$.
For  $\Omega_{\mathrm{m}} \geq 1$,
there are no possible scale-invariant models, consistent with causality relations in an expanding Universe \citep{MaedGueor21}.

While   $t$  varies  between $t_{\mathrm{in}}$  and
$t_0=1$ at  present,
the usual timescale $\tau$  in years or seconds   varies  from $\tau_{\mathrm{in }}=0$   and
$\tau_0= 13.8$ Gyr at  present  \citep{Frie08}.
The  relation between these  age units    is
$\frac{\tau - \tau_{\mathrm{in}}}{\tau_0 - \tau_{\mathrm{in}}} = \frac{t - t_{\mathrm{in}}}{t_0 - t_{\mathrm{in}}}$,
expressing that the age fractions of an event  are the same.
This~gives\vspace{-6pt}
\begin{eqnarray}
\tau \,= \, \tau_0 \, \frac{t- t_{\mathrm{in}}}{1- t_{\mathrm{in}}} \,  \quad \mathrm{and} \; \;
t \,= \, t_{\mathrm{in}} + \frac{\tau}{\tau_0} (1- t_{\mathrm{in}}) \,,\label{T2} \\
\mathrm{with} \quad \frac{d\tau}{dt} \, = \, \frac{\tau_0}{1-t_{\mathrm{in}}}\,, \quad \mathrm{and}\; \;
\frac{dt}{d\tau} \, = \, \frac{1-t_{\mathrm{in}}}{\tau_0}\,.   \label{dT1}
\end{eqnarray}

For increasing  $\Omega_{\mathrm{m}}$,  
the timescale $t$ is squeezed, since $t_{\mathrm{in}}=\Omega_{\mathrm{m}}^{1/3}$ is increasing,  
which reduces the range of $\lambda$ variations.
Both timescales are evidently linearly related.
Thus, for a given $\Omega_{\mathrm{m}}$, the derivatives are constant, 
which is useful for connecting the time variations, since   $\frac{dX}{d\tau}= \frac{dX}{dt} \frac{dt}{d\tau}$.
\subsection{Newton-like Dynamics}   \label{neuton}

The weak-field approximation is obtained from the above geodesics, with the following assumptions:
$g_{ij}=-\delta_{ij}$, $g_{00}= 1+\frac{\Phi}{c^2}$, $\Gamma^i_{00} =
\frac{1}{c^2} \partial^{i} \Phi$, $ \partial^i= g^{i \alpha} \frac{\partial}{\partial x^{\alpha}}$,
with $\Phi= -GM/r$ being the Newtonian potential. In addition, for slow motions,
$u^i \rightarrow  \frac{\upsilon^i} {c}$ and  $u^0 \rightarrow 1$. The geodesic Equation (\ref{geod}) becomes
\begin{equation}
\frac{1}{c^2} \frac{d \upsilon^i}{dt}+\frac{1}{c^2} \partial^i \Phi -\kappa \frac{\upsilon^i}{c^2}= 0 \,,
\label{Nvecz}
\end{equation}
\noindent
where $\kappa = -\frac{\dot{\lambda}}{\lambda}= \frac{1}{t}$.
This leads  to  the modified Newton equation in $t$-scale 
\citep{MBouvier79,Maeder17c},
\begin{equation}
\frac {d^2 \bf{r}}{dt^2} \, = \, - \frac{ G_t \, M}{r^2} \, \frac{\bf{r}}{r}   +  \kappa(t) \,\frac{d\bf{r}}{dt} \, ,
\label{Nvec}
\end{equation}

Equation (\ref{conserv}) imposes  for a dust Universe the relation  $\varrho a^3 \lambda=const.$,  %
implying that the  mass of an object is a coscalar of power $\Pi(M)=1$, thus $M= M(t) = M(t_0) \frac{t}{t_0}$.
In current units, according to (\ref{T2}), this becomes
\begin{equation}
M(\tau) = M(\tau_0) (t_{\mathrm{in}}+\frac{\tau}{\tau_0}(1-t_{\mathrm{in}}).
\label{MMM}
\end{equation}

A variation in mass is also a common situation in Special Relativity, where masses change for velocities tending to $c$.
Noticeably, the  potential  $\Phi =G\, M/r$
of an object appears as a more fundamental quantity,  being  scale-invariant  throughout the evolution of the~Universe.

We emphasize that this (possibly surprising) mass variation is also consistent in scale invariance
with the  Lagrangian definition of mass in mechanics, where the mass
appears as a proportionality factor between the Lagrangian  $\mathcal{L}$
and $\upsilon^2$ the square of the   velocities \citep{Landau60}. As shown in Appendix \ref{appb}, the  Lagrangian definition, together with
the  action principle  in scale invariance and  the need for  preserving  the  uniformity of the space-time,
imposes that the masses are of power $\Pi=1$, thus leading to
a variation with time.

In a  non-empty Universe, the effects of mass changes are rather limited.
As an example,
for $\Omega_{\mathrm{m}} =0.3$, the mass  at the Big Bang was
$ M(t_{\mathrm{in}})= 0.6694 \; M(t_0)$, since $M(t_{\mathrm{in}}) =  \Omega_{\mathrm{m}}^{1/3} M(t_0) $.
Over the last 400 Myr the variations were smaller than 1\%. This relative ``constancy'' of the mass over long periods
is what has allowed us to show that MOND theory is a valid approximation of the present SIV theory over the rotation period
of spiral galaxies~\citep{Maeder23}.

Equation (\ref{Nvec}) is expressed in variable $t$, where the age of the Universe is  $(1- t_{\mathrm{in}})$.
One has $\frac{d^2 r}{dt^2} =\frac{d^2 r}{d \tau^2} (\frac{d\tau}{dt})^2$, and
the constant $G_t$   in $t$-units becomes
$G_t (\frac{dt}{d\tau})^2= G$  in current units. 
The above equation becomes at present 
$\tau_0$,
\begin{equation}
\frac {d^2 \bf{r}}{d \tau^2}  = - \frac{G  M }{r^2}  \frac{\bf{r}}{r}   + \frac{\psi_0}{\tau_0}   \frac{d\bf{r}}{d\tau} ,
\quad \mathrm{with} \; \; \psi_0 =1-t_{\mathrm{in}} \,.
\label{Nvec4}
\end{equation}

The  additional term on the right is an acceleration  in the direction of the motion, we call it {\it{{the dynamical gravity}
}},
which is   usually very small, since $\tau_0$ is very large.
This term, proportional to the velocity,  favors collapse during a contraction,   and produces  an outwards acceleration
in an expansion.  The parameter  $\psi_0$ only applies at  present.
For $\Omega_{\mathrm{m}} =0,  0.05, 0.10, 0.20, 0.30, 1$,  one has $\psi_0=1, 0.632, 0.536, 0.415, 0.331, 0.$
In other epochs $\tau$,  instead of $\psi_0$ in Equation (\ref{Nvec4}), one has the numerical factor $\psi$
\begin{equation}
\psi(\tau) = \frac{t_0-t_\mathrm{in}}{\left( t_\mathrm{in} +   \frac{\tau}{\tau_0} (t_0-t_\mathrm{in})\right) }
\quad \left(  = \frac{t_0-t_{\mathrm{in}}}{t}  \; \right).
\label{ps}
\end{equation}

We now study  the time $\tau$ variations in the $\psi$ in the current time units.
Let us express the derivative of the numerical factor $\psi$ given in Equation (\ref{ps}),
\begin{equation}
\frac{d \psi}{d\tau} = \frac{d \psi}{dt}\frac{dt}{d\tau}=- \frac{(1-t_{\mathrm{in}})^2}{t^2  {\tau_0}}= -\frac{\psi^2}{\tau_0},
\quad \mathrm{and}  \; \; \frac{\dot{\psi}}{\psi} = - \frac{\psi}{\tau_0}.
\label{dpsi2}
\end{equation}

Moreover, since one has $\kappa=-\dot{\lambda}/\lambda=1/t$,  this implies $\dot{\kappa}=-\kappa^2$.

\subsection{The Two-Body Problem Within the SIV Paradigm}

To fully appreciate the dynamical effects of scale invariance, it is appropriate to examine
the two-body problem \citep{MBouvier79,Maeder17c}.
In these references, the equations were given in the  timescale $t$. Here, we better give them in the $\tau$-scale (years, seconds).
The equation of motion (\ref{Nvec}) can be written in plane polar coordinates $r$ and $\varphi$
\begin{eqnarray}
\ddot{r} - r \, \dot{\varphi}^2 \, = \,- \frac{G \, M}{r^2} +\frac{\psi}{\tau_0}\, \dot{r} \, ,
\label{Nr} \\
r \, \ddot{\varphi} + 2 \, \dot{r} \, \dot{\varphi} \, = \, \frac{\psi}{\tau_0}\, r \, \dot{\varphi}\,.
\label{Ntheta}
\end{eqnarray}

The term $\kappa(t)$ in $t$-time is replaced by $\kappa(\tau)=\psi(\tau)/\tau_0$ in $\tau$-units.
The integration of this last equation gives the equivalent law of angular momentum conservation
\begin{equation}
\frac{\psi}{\tau_0}\, r^2 \dot{\varphi} \, = \, L = \mathrm{const.}
\label{angm}
\end{equation}
for $\frac{d\psi}{d\tau}$ see Equation (\ref{dpsi2}).
Here, and in the rest of the text, $\varphi$ is the angular coordinate; that is $\varphi\in\left[0,2\pi\right]$.
$L$ is a scale-invariant quantity. This means that $ r^2 \, \dot{\varphi}$ increases like $\tau_0/\psi$, {{i.e.,}}
linearly with $t$.
After some manipulations  using Equation (\ref{angm}) to develop Equation (\ref{Nr})~\citep{Maeder17c},
we obtain the equivalent of the Binet equation,
\begin{equation}
\frac{d^{2} u}{d \varphi^{2}}  + u \, = \, \frac{GM}{L^{2}}  \left(\frac{\psi}{\tau_0}\right)^2, \quad \mathrm{where} \; \; u= 1/r .
\label{Binet}
\end{equation}
with an additional parenthesis.
Its general solution is a conic
\begin{equation}
r  \, = \, \frac{r_c}{1 + e \cos\varphi}, \quad \mathrm{with} \; \;
r_c = \, \frac{L^2}{G \, M }     \left(\frac{\tau_0}{\psi}\right)^2 .
\label{sol}
\end{equation}
\noindent
$r_c$ is  a particular solution for $ \frac{d^{2} u}{d \varphi^{2}} =0$:  the radius of a circular orbit ($e=0$).
For an elliptical orbital, the expression of the semi-major axis $a$ is
\begin{equation}
a= \frac{r_c}{1-e^2}.
\label{semi-major}
\end{equation}

From (\ref{sol}), we verify that $a$, like $r_c$,  scales with  $t$ (since $\psi \sim 1/t$); this means, for example,  that elliptical orbits slightly spiral outwards, with a constant eccentricity.

According to (\ref{angm}), the orbital velocity  $\upsilon$ of a circular motion behaves as follows:
\begin{equation}
\upsilon^2\, = \,( r_c \, \dot{\varphi})^2 = \frac{L^2}{r^2_c} \, (\frac{\tau_0}{\psi})^2.
\label{ups}
\end{equation}

This is a scale invariant quantity, since $r_c \sim  t$ and $\psi \sim  1/t$.  Now, we can express $L^2$
with (\ref{sol}) and obatin the usual expression:
\begin{equation}
\upsilon^2 \, = \, \frac{G\, M}{r_c} \,,
\label{v2}
\end{equation}

The scaling of $M$ and $r_c$  with $t$ confirms the scale invariance $\upsilon^2$,
which remains constant during the orbital expansion. In a subtle interplay,
the tangential acceleration of the ``dynamical gravity''  exactly compensates
the usual slowing down due to  orbital expansion.
This   is  also consistent with the time increase in $r^2 \, \dot{\varphi}$  (\ref{angm}).

The constancy of $\upsilon^2$ during expansion compares with MOND, where in the deep-limit, the  orbital
velocity becomes  independent from radius \citep{Milgrom09}, a key point regarding  the flat rotation curves of galaxies.
This concordance is not surprising, since   MOND appears as an approximation
of the Newton-like Equation (\ref{Nvec4}), 
{where} the scale factor $\lambda$ may be considered constant \citep{Maeder23}; 
an acceptable approximation over a few $10^8$ years.

\subsection{Secular Variations of the Orbital Parameters}

We now study  the time  variations of the orbital parameters in the current time units $\tau$.
Let us recall the derivative of the numerical factor $\psi$ given in Equation (\ref{ps}).
Using (\ref{dpsi2}) and (\ref{MMM}), based on (\ref{ups}) and (\ref{v2}) above, the radius $r_c$ of the circular orbit behaves as follows:
\begin{equation}
r_c = \, \frac{L^2}{G \, M }     (\frac{\tau_0}{\psi})^2 =
\frac{L^2}{G \, M(\tau_0) }     \frac{\tau^2_0  (t_{\mathrm{in}}+\frac{\tau}{\tau_0} (1-t_{\mathrm{in}})}{(1-t_{\mathrm{in}})^2}.
\label{RC}
\end{equation}

With $\dot{r_c}=\frac{L^2 \, \tau_0}{GM(\tau_0)(1-t_{\mathrm{in}})}$,
the relative variation with time $\tau$ of the semi-major axis $a$ given by (\ref{semi-major})
of an  elliptical orbit  becomes
\begin{equation}
\frac{\dot{a}}{a} = \frac{\dot{r_c}}{r_c}=   \frac{\psi}{\tau_0}
\quad \text{at time}\; \tau_0:  \psi=\psi_0= 1-t_{\mathrm{in}}\,.
\label{aa}
\end{equation}

As an example, for the Earth--Moon system, the above relation would predict a lunar recession amounting to 0.92 cm/yr
\citep{MaedGueor22}.
All variable quantities that have a linear dependence on time $t$ (and $\tau$) have a relative variation
equal to $\frac{\psi}{\tau_0}$. This is the case for the  orbital radius or semi-major axis, the mass and the  rotation period $T$.
For the masses, we can obtain from Equations (\ref{MMM})  and (\ref{ps})
\begin{equation}
\frac{\dot{M}}{M} =   \frac{\psi}{\tau_0}.
\label{Mps}
\end{equation}

Let us check the orbital period $T$, which is equal to $T=\frac{2 \, \pi}{\dot{\varphi}}$.
From  $r_c$ given by (\ref{sol}) and  the conservation law (\ref{angm}), we have
\begin{equation}
T= \frac{2 \, \pi r^2_c \psi}{L \tau_0} =
\frac{2 \pi  \psi L r^2_c}{L^2 \tau_0}= \frac{2 \pi L r_c \tau_0}{G M \psi}.
\end{equation}

Since $M$ and $r_c$ vary in the same way, the period $T$ varies like $1/\psi$; thus, based on Equation (\ref{dpsi2}),
\begin{equation}
\frac{\dot{T}}{T} \, =\,  - \frac{\dot{\psi}}{\psi}\,   =  \, \frac{\psi}{\tau_0}.
\label{TT}
\end{equation}

Remarkably, the above quantities strictly conserve {Kepler's third} law. We have the law of
angular momentum conservation (\ref{angm})  and the expression of the radius $r_c$ (\ref{sol}). Substituting $L$
from the first into the second,
\begin{eqnarray}
\frac{\psi}{\tau_0}\, r^2_c  \frac{2 \pi}{T} \, = \, L ,  \quad \mathrm{and} \quad
r_c = \, \frac{L^2}{G \, M }     (\frac{\tau_0}{\psi})^2\, . \nonumber \\
\mathrm{we\; get} \quad \frac{4 \pi^2 \,r^3_c}{GM \, T^2}= 1\, ,\quad \quad  \quad
\end{eqnarray}
where we considered circular orbits of  a test particle of negligible mass, to simplify the derivation. 
If it has a significant mass $m$, this has to be added to $M$.
The quantities  $T$, $M$, and $r_c$ all  have
the same functional dependence on time $\tau$, as illustrated by  the same relative variations
of the function $ \frac{\psi}{\tau_0}$. Thus,
we have a cubic dependence  on the same  function,  both in the numerator
and denominator, implying that at any time $\tau$ {Kepler's third} law is maintained.

\section{Conclusions}
\label{sec5}
We have revisited the scale-invariant field equations and some of the basic applications of the corresponding geodesic equations.
Along the way, we have demonstrated that any scale invariant matter action, 
or even a weaker symmetry known as  reparametrization invariance, 
naturally results in a ``standard'' energy--momentum tensor for matter of power 
$\Pi(\rho_m)=-2$ 
that can be related to a scale-invariant energy--momentum tensor upon 
augmenting the integration measure $\sqrt{-g}d^4x$ by the factor $\lambda^2$.
The example of a familiar ideal fluid was considered, to show the scaling power for energy-density and pressure.
The choice of $\lambda$ also makes specific predictions for an additional dynamic gravity term in the geodesics equations.
As such, this could be observationally accessed, and therefore this is not just another choice of gauge that does not have 
observational effects. It may be just the thing we need to get out of the current dark ages, dominated by dark matter and dark energy, 
and into enlightenment, where the scale invariance and/or reparametrization invariance could be the answer to understanding
our Universe without  a need for dark stuff.

The justification using an action principle of the scale invariant field equations, of the geodesics equations, 
and of the gauging condition for the scale factor $\lambda$ 
strengthens the theoretical basis of the scale-invariant paradigm, 
for which several observational tests have provided positive support for the theory.  
The equivalent Newton-like equation, the two-body problem, 
its secular variations, and the modified angular momentum conservation are given in conventional time units in the current epoch. 
The constancy of the orbital velocity during secular expansion is an interesting consequence
in relation with the flat rotation curves of galaxies and  similar problems in very wide binaries \citep{Hernandez22}.
Kepler's third law remains a rock untouched by scale-invariance effects.

\vspace{6pt}

\authorcontributions{
Conceptualization, A.M.; 
Formal analysis, A.M. and V.G.G.; 
Investigation, A.M. and V.G.G.; 
Methodology, A.M.; 
Validation, A.M. and V.G.G.; 
Writing---original draft, A.M.; 
Writing---review and editing, A.M. and V.G.G.
Both co-authors were actively involved in the writing of the paper and its draft versions. 
All authors have read and agreed to the published version of the manuscript.}

\funding{This research received no external funding.}

\dataavailability{No new data were created or analyzed in this study.} 


\acknowledgments{A.M. expresses his gratitude to his wife for her patience and support. 
V.G.G. is extremely grateful to his wife and daughters for their understanding and family support  during the various stages of the research presented. 
This research did not receive any specific grant from funding agencies in the public, commercial, or not-for-profit sectors.}

\conflictsofinterest{The authors declare no conflict of interest.} 

\appendixtitles{yes} 
\appendixstart
\appendix
\section[\appendixname~\thesection]{Detailed Calculations Using the Action  Principle}\label{appa}


{We feel that it is important to provide a concise derivation of the main equations 
within the body of our paper, instead of referring to other papers and the derivations there, 
which could use different symbols and conventions. Thus, here it is.}

From Equation (\ref{ig2}), the action principle for scale invariance symmetry is written \citep{Dirac73},
\begin{equation}
\delta I_{\mathrm{G}} \, = \,\delta   \int  \, \left( -\lambda^2 \, R  + 6 \, \lambda^{\mu} \lambda_{\mu}
+ c_2 \, \lambda^4  \right) \sqrt{-g} \, d^4 x \,,
\label{ig22}
\end{equation}

The variation in the second term is
\begin{equation}
\delta( \lambda^{\mu} \lambda_{\mu} \sqrt{-g})= \lambda^{\mu} \sqrt{-g}\, \delta \lambda_{\mu}+
\lambda_{\mu} \sqrt{-g} \delta \lambda^{\mu}+\lambda^{\mu} \lambda_{\mu} \delta \sqrt{-g}\,,
\label{Ea}
\end{equation}

The  $2\text{nd}$ term on the r.h.s. is $\lambda_{\mu} \sqrt{-g} \delta \lambda^{\mu}=
\lambda_{\mu} \sqrt{-g} \delta ( g^{\mu \nu } \lambda_{\nu})$; thus, we develop \mbox{Equation (\ref{Ea})} into
\begin{eqnarray}
\delta( \lambda^{\mu} \lambda_{\mu} \sqrt{-g})= 2 \lambda^{\nu} \sqrt{-g}\, \delta \lambda_{\nu}+
\lambda_{\mu} \lambda_{\nu} \sqrt{-g} \, \delta g^{\mu \nu} +\lambda^{\mu} \lambda_{\mu} \delta \sqrt{-g}\,,
\label{e1}\\
= -2 (\lambda^{\nu}\sqrt{-g})_{,\nu} \delta \lambda - \lambda^{\mu} \lambda^{\nu}  \sqrt{-g}\,\delta g_{\mu \nu}+
\frac{1}{2} \lambda^{\alpha} \lambda_{\alpha} g^{\mu \nu}  \sqrt{-g}\,\delta g_{\mu \nu}\,. \label{e2}
\end{eqnarray}

The $1\text{st}$ term on the r.h.s. results from the manipulation of
$ (\lambda^{\alpha}\sqrt{-g})_{,\alpha} \delta \lambda +\lambda^{\alpha}\sqrt{-g} \,  \delta \lambda_{\alpha}
$ = $  (\lambda^{\alpha} \sqrt{-g}\,\delta\lambda)_{,\alpha}$ 
being  an exact differential, which can thus be eliminated \citep{Landau60}.
The $2\text{nd}$ comes from $g_{\mu \nu} dg^{\mu \nu}= -g^{\mu \nu} dg_{\mu \nu}$, 
and the $3\text{rd}$  from $\delta \sqrt{-g}=\frac{1}{2} \sqrt{-g}\, g^{\mu \nu}\delta g_{\mu \nu}$ (e.g., \mbox{Section 4.7} \citet{Weinberg78}).
We note that a sequence of developments similar  to that  of $\delta( \lambda^{\mu} \lambda_{\mu} \sqrt{-g})$ is already
present in Section 90 of the book by \citet{Eddington23}.
Using the remark after (\ref{ez}), we have
\begin{equation}
\delta(\lambda^{\alpha}\lambda_{\alpha} \sqrt{-g})=- 2\lambda^{\alpha}_{;\alpha}\sqrt{-g} \delta \lambda
- \left(\lambda^{\mu} \lambda^{\nu} -\frac{1}{2} g^{\mu \nu} \lambda^{\alpha} \lambda_{\alpha}\right)
\sqrt{-g}\,\delta g_{\mu \nu}\, . \label{e3}
\end{equation}

Now, we examine  $\lambda^2 \,  \sqrt{-g} \,R$ in the action (\ref{ig2}).
First, the variation in $R= g^{\mu \nu}  R_{\mu \nu}$ is:
\begin{equation}
\delta( \sqrt{-g} R)= R_{\mu \nu} \delta(g^{\mu \nu}\sqrt{-g}) +\sqrt{-g}( g^{\mu \nu} \delta \Gamma^{\sigma}_{\nu \sigma})_{; \mu}  -
\sqrt{-g}(g^{\mu \nu}\delta \Gamma^{\rho}_{\mu \nu} ) _{; \rho}  \,.
\end{equation}

Since  $\delta R_{\mu \nu}= (\delta  \Gamma^{\sigma}_{\mu \sigma})_{; \nu} - (\delta  \Gamma^{\sigma}_{\mu \nu})_{; \sigma} $,
see \citet{Weinberg78}  Section 12.4.
Account is also given that $(g^{\mu \nu})_{; \alpha}=0$ and
$A^{\mu}_{; \,\mu}= \frac{1}{\sqrt{-g}} \frac{\partial }{\partial x^{\mu}} \sqrt{-g} A^{\mu}$; thus,
we  have
\begin{equation}
\delta( \sqrt{-g} R) \, =  R_{\mu \nu} \delta(g^{\mu \nu}\sqrt{-g})+
\left( g^{\mu \nu} \sqrt{-g}\delta   (g^{\rho}_{\mu} \Gamma^{\sigma}_{\nu \sigma}-
\Gamma^{\rho}_{\mu \nu} )\right)_{, \rho}  \, .
\label{f1}
\end{equation}

The  derivatives of both $\Gamma$-terms are now expressed
with respect to the same coordinate  $\rho$, since $g^{\rho}_{\mu} (\Gamma^{\sigma}_{\nu \sigma})_{, \rho} =
(\Gamma^{\sigma}_{\nu \sigma}) _{ , \mu}$.  We now include the term $\lambda^2$.
The term originating from $\delta R_{\mu \nu}$ vanishes
in standard theory, while here  it has some contribution to the action \citep{Dirac73}. The first term in the action gives
\begin{eqnarray}
\delta (\lambda^2\sqrt{-g} R) \, =   \lambda^2 R_{\mu \nu} \delta (g^{\mu \nu} \sqrt{-g})+ \nonumber \\
2 \lambda \lambda_{\rho}  g^{\mu \nu} \sqrt{-g}\delta \left(  \Gamma^{\rho}_{\mu \nu} -
g^{\rho}_{\mu} \Gamma^{\sigma}_{\nu \sigma}  \right)
+2 \lambda R \sqrt{-g} \delta \lambda\,, \label{f2}
\end{eqnarray}
since we have the following  exact  differential, which vanishes after the integration.
\begin{eqnarray}
(\lambda^2 g^{\mu \nu} \sqrt{-g}\delta R_{\mu \nu})_{,\rho}=
\lambda^2 \left( g^{\mu \nu} \sqrt{-g}\delta (  g^{\rho}_{\mu} \Gamma^{\sigma}_{\nu \sigma} -
\Gamma^{\rho}_{\mu \nu} )\right) _{, \rho} - \nonumber \\
2 \lambda \lambda_{\rho}  g^{\mu \nu} \sqrt{-g}\delta \left(  \Gamma^{\rho}_{\mu \nu} -
g^{\rho}_{\mu} \Gamma^{\sigma}_{\nu \sigma}  \right)\,.
\end{eqnarray}

\textls[-15]{Let us examine  the first term of the r.h.s  in Equation (\ref{f2}), we use the following {relations}}:
\begin{eqnarray}
R_{\mu \nu} \delta(g^{\mu \nu} \sqrt{-g}) =  R \delta\sqrt{-g}+R_{\mu \nu} \sqrt{-g} \delta g^{\mu \nu}, \\
\delta \sqrt{-g}= \frac{1}{2} \sqrt{-g} g^{\mu \nu} \delta g_{\mu \nu}, \quad \mathrm{and} \; \;
\delta g^{\mu \nu}= -g^{\mu \rho} g^{\nu \sigma} \delta  g_{\rho \sigma}, \label{detg}\\
\mathrm{thus} \; \;    \lambda^2 R_ {\mu \nu} \delta  (g^{\mu \nu} \sqrt{-g}) =
\lambda^2 \left(\frac{1}{2} g^{\mu \nu} R - R^{\mu \nu} \right) \sqrt{-g} \delta g_{\mu \nu } \,.
\label{f3}
\end{eqnarray}

Equation  (\ref{f2}) becomes, with  a reformulation of the second term on the r.h.s. \citep{Dirac73},
\begin{eqnarray}
\delta (\lambda^2\sqrt{-g} R) \, =  \lambda^2 \left(\frac{1}{2} g^{\mu \nu} R -
R^{\mu \nu} \right) \sqrt{-g} \delta g_{\mu \nu }+\nonumber \\
2  g^{\mu \nu}(\lambda \lambda^{\rho})_{; \rho} \sqrt{-g} \delta g_{\mu \nu} -2 (\lambda \lambda^{\mu})^{; \nu}
\sqrt{-g} \delta g_{\mu \nu} + 2 \lambda R \sqrt{-g} \delta \lambda\,, \label{f4}
\end{eqnarray}

Finally, the last term in the action (\ref{ig2}) gives
\begin{equation}
\delta (c_2 \lambda^4  \sqrt{-g}) = 4 c_2 \lambda^3\sqrt{-g} \delta \lambda +
\frac{1}{2} c_2 \lambda^4 g^{\mu \nu} \sqrt{-g} \delta g_{\mu \nu} \, ,
\label{f5}
\end{equation}
where we have used  the first relation in Equation (\ref{detg}).


We now have the various contributions to the action (\ref{ig2}).
In order to have an extremum   with $\delta I_{\mathrm{G}}=0$, the parenthesis in Equation (\ref{ig2}) must vanish.
From relations (\ref{e3}), (\ref{f4}), and (\ref{f5}), we obtain the terms contributing $\delta g_{\mu \nu}$,
\begin{eqnarray}
-6 \lambda^{\mu} \lambda^{\nu}+3 g^{\mu \nu} \lambda^{\alpha} \lambda_{\alpha}+
\lambda^2  (R_{\mu \nu} - \frac{1}{2} g^{\mu \nu} R ) \nonumber \\
- 2 g^{\mu \nu}(\lambda \lambda^{\rho}){_{; \rho}}+
2 (\lambda \lambda^{\mu}){^{; \nu}}+ \frac{1}{2}  c_2 \lambda^4 g^{\mu \nu} =0 \, .
\label{f6}
\end{eqnarray}

The above equation represents  the scale invariant field equation for the vacuum with a  curvature \citep{Dirac73}
associated with the cosmological constant, like in the de Sitter model.
Aside from the missing mass term $g_{\mu\nu}\mu_0\lambda^2$, the above expression is practically the same
modified energy--momentum tensor as discussed by \citet{Parker'73}.
We  perform the following identification
$c_2  = 2\Lambda_{\mathrm{E}}$, 
where $\Lambda_{\mathrm{E}}$ is the cosmological constant in GR 
and $\Lambda= \lambda^2 \Lambda_{\mathrm{E}}$ is
the value in SIV theory.
The  dependence in
$\lambda^2$ is consistent with that of a density, see Equation (\ref{ro2}).

After division by $\lambda^2$ and  regrouping of the terms in (\ref{f6}), one can write
\begin{eqnarray}
R^{\mu \nu} - \frac{1}{2} g^{\mu \nu} R - 2 g^{\mu \nu} \frac{(\lambda^{\rho}){_{; \rho}}}{\lambda}
+ 2 \frac{ ( \lambda^{\mu}){^{; \nu}}}{\lambda} \nonumber \\
+g^{\mu \nu} \frac{\lambda^{\alpha} \lambda_{\alpha}}{\lambda^2}   - 4 \frac{\lambda^{\mu} \lambda^{\nu}}{\lambda^2}+
\lambda^2 \Lambda_{\mathrm{E}} \,g^{\mu \nu} =0 \, ,
\label{f6x}
\end{eqnarray}

The above relation represents the scale-covariant field equation for the empty space with a cosmological constant,
it is also  the first member of scale-invariant field equation when matter is present.

\section{The Mass Scaling and the Lagrangian for Matter}
\label{appb}
The invariance of $T_{\mu \nu}$  in Equation (\ref{pr2}) imposes that densities (and pressures)  vary like
$\varrho = \varrho' \, \lambda^2$
according to Equation (\ref{ro2}). This means that, for a mass $M$ in a volume $V$,
\begin{equation}
\frac{M}{V}\, = \; \frac{M'}{V'} \lambda^2, \quad \mathrm{with} \; V'= \lambda^3\, V  \quad  \Rightarrow M'=\lambda M .
\label{m1}
\end{equation}

Thus, the invariance of $T_{\mu \nu}$ imposes that the masses are of power $\Pi=1$; {{i.e.,}} that the  masses behave
like  $t$.

Let us consider  a free test particle  moving in empty space in classical mechanics.
Instead of an action integrated over  four-coordinate space-time, we consider an action integrated over time
with $\sqrt{-g} = 1$ \citep{Landau60},
\begin{equation}
\delta I \, = \,\delta \int  \lambda^2 \mathcal{L}(q, \dot{q}, t) \, dt \,= \, 0\,. \quad   \quad {q: \; coordinate},
\label{lagr}
\end{equation}
with a factor $\lambda^2$ as in Equation (\ref{final}), since the Lagrangian is proportional to energy density.
The uniformity of space-time implies that the Lagrangian $\mathcal{L}(q, \dot{q}, t)$
cannot  depend on the location $q$ or time $t$;
thus, it  can only depend on  velocity {\textbf{$\upsilon$}}.
However, the isotropy of space excludes
a dependence on vector {\textbf{$\upsilon$}}; thus, only a dependence on the module $\upsilon^2$ is possible; 
thus, one possibility is the familiar quadratic expression: 
$ \mathcal{L} \, = \, a \, \upsilon^2$ with $a$ as a proportionality factor.
For small variations in the velocity $\delta\upsilon$, the  Lagrangian becomes
\begin{eqnarray}
\mathcal{L}(\upsilon+\delta\upsilon) = a \,(\upsilon+\delta\upsilon)^2 =
a \, \upsilon^2+ 2 \, a \upsilon \delta\upsilon+ a \, (\delta\upsilon)^2 \,,  \\
\mathcal{L}(\upsilon+\delta\upsilon) \, = \,\mathcal{L}(\upsilon) + \frac{d}{dt} (2\, a\, r \, \delta \upsilon + a (\delta\upsilon)^2 \,t) \,.
\end{eqnarray}

The 2$\text{nd}$ term on the right is a total derivative, which can be eliminated from the integral (cf. Appendix \ref{appa}),
so that the Lagrangian in Equation (\ref{lagr}) becomes \citep{Landau60},
\begin{equation}
\mathcal{L}= a \, \upsilon^2\,,
\end{equation}

Such an expression of $\mathcal{L}$ is  currently used in simple applications of the action principle in the standard case, as well as
being applicable in the case of scale invariance.
Now, we examine what this  form  of the Lagrangian  in the action principle also implies.
The factor $a$ is generally designated using  $m/2$, where $m$ is the particle mass.
Time is of power $\Pi=1$ and  $\lambda$  of power $-$1.
To make the invariance of the  action, $\mathcal{L}$
must be of power $\Pi=+1$, meaning  that $\mathcal{L}$  behaves like $t$, since $\upsilon^2$ is an invariant.
Thus, the factor $a=m/2$ and, therefore, the mass is of power $\Pi=+1$.
The condition of the scale invariance of the action together
with that of a uniform space-time imposes a power + 1 to the mass.
This result is in full agreement with the consequences of the conservation law as expressed, 
for example, by Equation (\ref{conserv}).

\label{lastpage}

\begin{adjustwidth}{-\extralength}{0cm}

\reftitle{References}

\PublishersNote{}
\end{adjustwidth}

\end{document}